\documentstyle[aps,prl,twocolumn]{revtex}   

\begin{document}

\draft

\wideabs{

\title{Quantum noise-induced chaotic oscillations}

\author{Bidhan Chandra Bag and Deb Shankar Ray}

\address{Indian Association for the Cultivation of Science,
Jadavpur, Calcutta 700 032, INDIA.}

\maketitle

\begin{abstract}
We examine the weak quantum noise limit of Wigner  equation for phase 
space distribution functions. It has been shown that the leading order 
quantum noise described in terms of an auxiliary Hamiltonian manifests 
itself as an additional fluctuational degree of freedom which may induce 
chaotic and regular oscillations in a nonlinear oscillator. 
\end{abstract} 


}

The absence of any direct counterpart to classical trajectories in phase 
space in quantum theory poses a special problem in nonlinear dynamical
system from the point of view of quantum-classical correspondence 
\cite{gutz,stev,Heller}. As an
essential step towards understanding quantum systems a number of semiquantum 
methods, via WKB approximation, Ehrenfest theorem or mean field approximation
as well as some exact calculations etc. have been proposed and investigated over the years
\cite{gutz,stev,Heller,ash,patt,schanz,milonni,zurek}. 
A particularly noteworthy case \cite{ash} concerns a system that seems to be classically
integrable but not in the quantum case due to tunneling.
In the present paper
we examine a related issue, i. e, the weak quantum noise limit of Wigner
equation for phase space distribution functions and show that it is possible
to describe the quantum fluctuations of the system in terms of an auxiliary
degree of freedom within an effective Hamiltonian formalism. This allows us to 
demonstrate an interesting quantum noise-induced chaotic and 
regular behaviour in a driven double-well oscillator.

To start with we consider a one-degree-of-freedom system described by the 
Hamiltonian equation of motion ;

\begin{eqnarray}
\dot{x} & = & \frac{\partial H}{\partial p} \; \; = p \nonumber\\
\dot{p} & = & - \frac{\partial H}{ \partial x} \; \; = -V'(x, t)
\end{eqnarray}

where $x$ and $p$ are the co-ordinate and momentum variables for the system
described by the Hamiltonian $H (x, p, t)$. $V(x, t)$ refers to the potential
of the system. The reversible Liouville dynamics corresponding to Eq.(1) is given
by

\begin{equation}
\frac{\partial \rho}{\partial t} = -p \frac{\partial \rho}{\partial x} +
V'(x, t)\frac{\partial \rho}{\partial p} 
\end{equation}

Here $\rho(x, p, t)$ is the classical phase space distribution function. For a 
quantum-mechanical system, however, $x$, $p$ are not simultaneous observables
because they become operators which obey Heisenberg uncertainty relation.
The quantum analog of classical phase space distribution function $\rho$ 
corresponds to Wigner phase space function $W(x, p, t)$ ;  
$x$, $p$
now being the c-number variables. $W$ is given by Wigner equation \cite{wigner};

\begin{eqnarray}
\frac{\partial W}{\partial t} & = &
-p \frac{\partial W}{\partial x}
+ V'(x, t) \frac{\partial W}{\partial p} \nonumber \\
& & + \sum_{n\geq 1} 
\frac{ \hbar^{2n}(-1)^n}{ 2^{2n}(2n+1)! }
\frac{\partial^{2n+ 1}V}{\partial x^{2n+ 1} }
\frac{ \partial^{2n+1} W }{\partial p^{2n+1} }  \; \; .
\end{eqnarray} 

The third term in Eq.(3) corresponds to quantum correction to classical
Liouville dynamics.

Our aim in this report is to explore an auxiliary Hamiltonian description 
corresponding to Eq.(3) in the semiclassical limit $\hbar \rightarrow 0$.
To put this in an appropriate context let us bring forth below an analogy with an
observation \cite{nature} on a weak thermal noise limit of overdamped  
Brownian motion of a particle in a force field.

In that significant analysis, Luchinsky and McClintock \cite{nature} 
have studied the
large fluctuations (of the order $>> \sqrt D, D$ being the diffusion coefficient)
of the dynamical variables $\vec x$ away from and return to the stable 
state of the system with
a clear demonstration of detailed balance. 
The physical situation is governed
by the standard Fokker-Planck equation for 
probability density $P_c(\vec x, t)$,

\begin{equation}
\frac{\partial P_c (\vec x, t)}{\partial t} = - \vec \nabla \cdot \vec K(\vec x, t)
P_c(\vec x, t) + \frac{D}{2} \nabla^2 P_c(\vec x, t) \; \; \; ,
\end{equation}

where $\vec K(\vec x, t)$ denotes the force field.

In the weak noise limit $D$ is considered to be a 
smallness parameter such that in the
limit $D \rightarrow$ small, $P_c(\vec x, t)$ can be described by a WKB-type  
approximation of the Fokker-Planck equation \cite{nature,tel} 
of the form $P_c(\vec x, t) =
z(\vec x, t) \exp (\frac{w (\vec x, t)}{D})$ . Here $z(\vec x, t)$ is a prefactor
and $w(\vec x, t)$ is the classical action satisfying the Hamilton-Jacobi 
equation which can be solved by integration of an auxiliary Hamiltonian equation 
of motion \cite{nature}
\begin{eqnarray}
\dot {\vec x} & = & 
\vec p + \vec K \; \; ,  \; \; \dot {\vec p} = -\frac{\partial \vec K}
{\partial \vec x} \vec p \nonumber\\ 
H_{aux}(\vec x, \vec p, t) & = & \vec p \cdot \vec K(\vec x, t) 
+ \frac{1}{2} \vec p \cdot \vec p \; \; ,
\; \; \vec p = \vec \nabla w \; \; ,
\end{eqnarray}
where $\vec p$ is a momentum of the auxiliary system.

The origin of this auxiliary momentum $\vec p$ is the fluctuations of the 
reservoir. In a thermally equilibrated system as emphasized by Luchinsky 
and McClintock \cite{nature}, a typical large fluctuation of the variable  $\vec x$ implies
a temporary departure from its stable state $\vec x_s$ to some remote state
$\vec x_f$ (in presence of $\vec p$) followed by a return to $\vec x_s$ as a
result of relaxation in the absence of fluctuations $\vec p$ (i. e. , $\vec p = 0$).
Luchinsky and McClintock have studied these fluctuational and relaxational
paths in analog electronic circuits and demonstrated the symmetry of growth and 
decay of classical fluctuations in equilibrium.

We now return to the present problem and in analogy to weak thermal noise limit
we look for the weak quantum noise limit of Eq.(3) by setting $\hbar \rightarrow 0$
with $W(x, p, t)$ described by a WKB type approximation of the form

\begin{equation}
W(x, p, t) = W_0(x, t) \exp(-\frac{s(x, p, t)}{\hbar}) \; \;.
\end{equation}

where $W_0$ is again a pre-exponential factor and $s(x, p, t)$ is the 
classical action function satisfying  Hamilton-Jacobi equation which can be
solved by integrating the following Hamilton's equations

\begin{eqnarray}
\dot{x} & = & p \nonumber\\
\dot{X} & = & P \nonumber\\
\dot{p} & = & V'(x, t) -
\sum_{n\geq 1}  \frac{(-1)^{3n+1}}{2^{2n}}\frac{1}{(2n)!} 
\frac{\partial^{2n+1} V}{\partial x^{2n+1}} X^{2n} \nonumber\\
\dot{P} & = & V''(x, t) X -
\sum_{n\geq 1} \frac{(-1)^{3n+1}}{2^{2n}(2n+1)!} 
 \frac{\partial^{2(n+1)} V}{\partial x^{2(n+1)} } X^{2n+1}
\end{eqnarray}

with the auxiliary  Hamiltonian $H_{aux}$

\begin{equation}
H_{aux}  =  p P - V'(x,t) X 
+ \sum_{n\geq 1} \frac{(-1)^{3n+1}X^{2n+1}}{2^{2n}(2n+1)!} 
\frac{\partial^{2n+1} V}{\partial x^{2n+1}} 
\end{equation} 
where we have defined the auxiliary co-ordinate $X$ and momentum $P$ as
\begin{equation}
X = \frac{\partial s}{\partial p} \; \; \; {\rm and} 
\; \; P = \frac{\partial s}{\partial x} \; \;.
\end{equation} 

The interpretation of the auxiliary variables $X$ and $P$ is now derivable
from the analysis of Luchinsky and McClintock \cite{nature}.
The introduction of $X$ and $P$ in the dynamics implies the addition of a new 
degree of freedom into the classical system originally described by $x, p$. Since the 
auxiliary degree of freedom ($X, P$) owes its existence to the weak quantum 
noise, we must look for the influence of weak quantum fluctuations on the
dynamics in the limit $X \rightarrow 0$, $P \rightarrow 0$, so that the 
Hamiltonian tends to be vanishing (since the $X$ and $P$ appear as
multiplicative factors in the auxiliary Hamiltonian $H_{aux}$).
It is therefore plausible that this vanishing Hamiltonian method captures 
the essential features of some generic quantum effect of the dynamics in 
classical terms in the weak quantum fluctuation limit. In what follows we shall be
concerned with a quantum noise-induced barrier crossing dynamics - as a typical
effect of this kind in a driven double-well system. Furthermore since the 
auxiliary Hamiltonian describes an effective two-degree-of-freedom system, the 
system, in general, by virtue of nonintegrability may admit chaotic behaviour. This
allows us to study a dynamical system where one of the degrees of freedom is of
quantum origin. Thus if the driven one degree-of-freedom is chaotic, the
influence of the quantum fluctuational degree of freedom on it appears to be
quite significant from the point of view of what may be termed as quantum chaos.
We point out, in passing, that the Wigner function approach 
of somewhat different kind, has also been considered earlier by 
Zurek and others \cite{zurek} for the analysis of quantum decoherence problem
in the context of quantum-classical correspondence.

The testing ground of the above analysis is a driven double
well oscillator characterized by the following Hamiltonian
\begin{eqnarray}
H & = & \frac{p^2}{2} + V(x, t) \; \;, \nonumber\\
V(x, t) & =  & ax^4 - b x^2 + g x cos \Omega t
\end{eqnarray}

where $a$ and $b$ are the constants defining the potential. $g$ includes the effect
of coupling with the oscillator with the external field with frequency $\Omega$.
The model described by (10) has been the standard paradigm for studying
chaotic dynamics over the last few  years \cite{sc1,sc2,bag1,lin}.

The equation of motion corresponding to auxiliary Hamiltonian $H_{aux}$ is
given by

\begin{eqnarray}
\dot{x} & = & p \nonumber\\ 
\dot{X} & = & P \nonumber\\
\dot{p} & = & 4 a x^3 -2 b x + g \cos{\Omega}t -3 a x X^2  \nonumber\\
\dot{P} & = & (12 a x^2 -2 b) X  - a X^3 
\end{eqnarray}

In order to make our numerical analysis that follows 
consistent with this scheme of weak quantum
noise limit it is necessary to consider limit of auxiliary Hamiltonian.
To this end we fix the initial condition for the quantum noise
degree of freedom $P = 0$ and Lt $X \rightarrow $ very small for the entire analysis. 
The relevant parameters for the numerical study \cite{bag1,lin} are $a = 0.5$,
$b =10$, $g = 10$ and $\Omega = 6.07$.

The results of numerical integration of Eq.(11) for the initial condition
of the oscillator $p =0$, $x = -2.512$ (along with $P = 0$ and $ X =1.5 \times
10^{-6}$) are shown in the Poincare plot (Fig. 1). What is apparent 
from a detailed follow-up of the system is that the system rapidly jumps
back and forth between the two wells at irregular intervals of time resulting in a
chaotic Poincare map spreaded over the two wells. This is in sharp contrast
to what we observe in Fig. 2 on plotting the results of numerical integration of
classical equations of motion corresponding to Eq.(1) and Hamiltonian
(10) with the same initial condition $p =0$ and $x = -2.512$. The system in this
case resides in the four islands of the left well. It is thus immediately
apparent that the quantum noise degree of freedom which imparts weak quantum
fluctuations in the system through very small but nonzero $X$ induces a passage
from left to right well and back.

In Fig.3 we fix the initial condition at a different turning point $p =0$, 
$x = -2.509$ and calculate the auxiliary Hamiltonian dynamics Eq.(11). It is interesting
to observe that the noise strength is not sufficient to make the system move from the 
left well where it stays permanently by depicting a closed regular curve on the
Poincare section.

The quantum noise-induced barrier crossing dynamics from left to right well
and back is illustrated in Figs.4(a-c). The initial condition for the oscillator
used in this case is $p =0$, $x = -2.5093$. The closed curve in Fig. 4(a) exhibits
a snapshot of the confinement of the system (in the left well) upto the time
$t = n T$ where $n = 1293$ and $T$ is the time period of the external field
($T = \frac{2 \pi}{\Omega}$). The system then jumps to the right well to stay there
for a period of time  2998 T. This is shown in Fig.4(b). The process goes on
repeating for the next period of time $2969 T$ when the system gets confined in 
the left well again. The back and forth quantum noise-induced oscillations
between the two wells illustrate a regular dynamics in this case. In the
absence of noise the classical system [Eq.(1)] remains localized in a specific
well.

In summary, we have shown that the leading order quantum noise in Wigner equation 
for phase space distribution functions results in an auxiliary Hamiltonian where the
quantum noise manifests itself as an extra fluctuational degree of freedom. 
Depending on the initial conditions this may induce irregular or regular
hopping between the two wells of a double-well oscillator. It is thus possible
that a nonlinear system may sustain chaotic oscillations by quantum noise, even
when its classical counterpart is fully regular.

\acknowledgments
B. C. Bag is indebted to the Council of Scientific and
Industrial Research (C.S.I.R.), Govt. of India, for partial financial support.

\begin{center}
{\bf Figure Captions}
\end{center}

\begin{enumerate}
\item Plot of $x$ vs $p$ on the Poincare surface of section ($X= 0$) for Eq.(13) with initial
condition $x = -2.512 $, $p = 0$, $X \rightarrow 0$, $P = 0$. (Units are arbitrary).
\item Plot of $x$ vs $p$ for Eq.1 with Hamiltonian (10) and initial condition
$x = -2.512$ and $p = 0.0 $.
\item Same as in Fig.1 but for $x = -2.509$ and $p = 0.0$.
\item Same as in Fig.1 but for $x =-2.5093$, and $p = 0$. The observations are taken for 
the time intervals $(a) t = 0 \; {\rm to} \; 1293 T$ (left well), 
$(b) t = 1293 T \; {\rm to} \; 4291 T$ (right well) and 
$(c) t = 4291 T {\rm to} \; 7260 T$ (left well).
[T (=$\frac{2 \pi}{\Omega}$) is the time period of the driving field].
\end{enumerate}


\begin{thebibliography}{99}

\bibitem{gutz}
M. Gutzwiller, {\it Chaos in Classical and Quantum Mechanics} 
(Springer, Berlin, 1990);
I. C. Percival, Adv. Chem. Phys. {\bf 36} 1 (1977).

\bibitem{stev}
P. Stevenson, Phys. Rev. {\bf D 30} 1712 (1984).

\bibitem{Heller}
E. J. Heller; in {\it Chaos and Quantum Physics} 
(Proceedings  of the Les Houches Summer School, 1989)
(North-Holland Amsterdam, 1991).

\bibitem{ash}
Y. Ashkenazy, L. P. Horwitz, J. Levitan, M. Lewkowicz and Y. Rothchild,
Phys. Rev. Letts. {\bf 75} 1070 (1995); Y. Ashkenazy, R. Berkovits, L. P.
Horwitz and J. Levitan, Physica {\bf A238} 279 (1997); L. P. Horwitz, 
J. Levitan and Y. Ashkenazy, Phys. Rev. {\bf E55} 3694 (1997).


\bibitem{patt}
A. K. Pattanayak and W. C. Schieve, Phys. Rev. Letts. {\bf 72} 2855 (1994).

\bibitem{schanz}
H. Schanz and B. Esser, Phys. Rev. {\bf A 55} 3375 (1997).

\bibitem{milonni} B. Sundaram and P. W. Milonni,
Phys. Rev. {\bf E 51} 1971 (1995).

\bibitem{zurek} 
W. H. Zurek and J. P. Paz, Phys. Rev. Lett. {\bf 72} 2508 (1994); S. Habib,
K. Shizume and W. H. Zurek, Phys. Rev. Lett. {\bf 80} 4361 (1998).

\bibitem{wigner} E. P. Wigner Phys. Rev. {\bf 40} 749 (1932). 

\bibitem{nature} 
D. G. Luchinsky and P. V. E. McClintock, Nature {\bf 389} 403 (1997).

\bibitem{tel} R. Graham and T. T\'el, Phys. Rev. Letts {\bf 52} 9 (1984).

\bibitem{sc1}
S. Chaudhuri, G. Gangopadhyay and D. S. Ray, Phys. Rev. E {\bf 47} 311 (1993). 

\bibitem{sc2}
S. Chaudhuri, G. Gangopadhyay and D. S. Ray, Phys. Rev. E {\bf 52} 2262 (1995). 

\bibitem{bag1} 
B. C. Bag, S. Chaudhuri, J. Ray Chaudhuri and D. S. Ray, Physica D {\bf 125} 
47 (1999).

\bibitem{lin} 
W. A. Lin and L. E. Ballentine, Phys. Rev. Lett. {\bf 65} 2927 (1990).

\end{thebibliography}
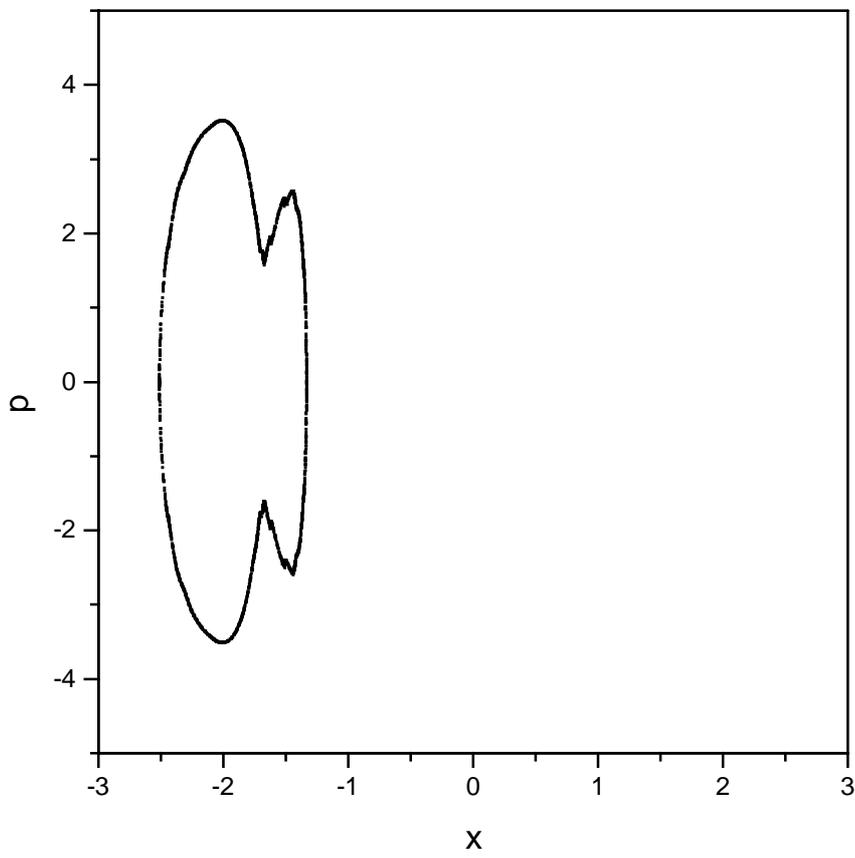
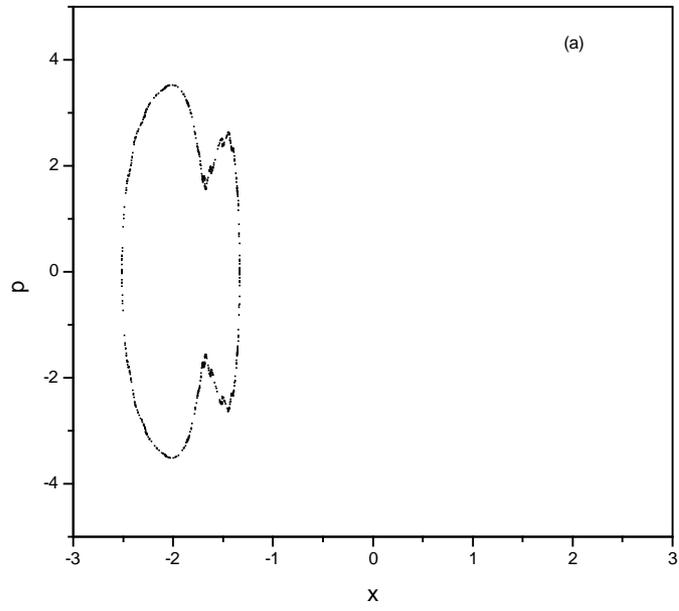
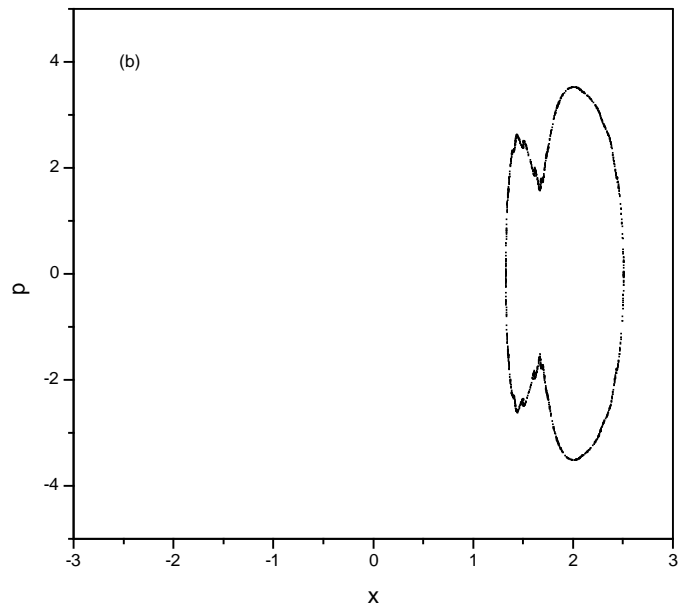
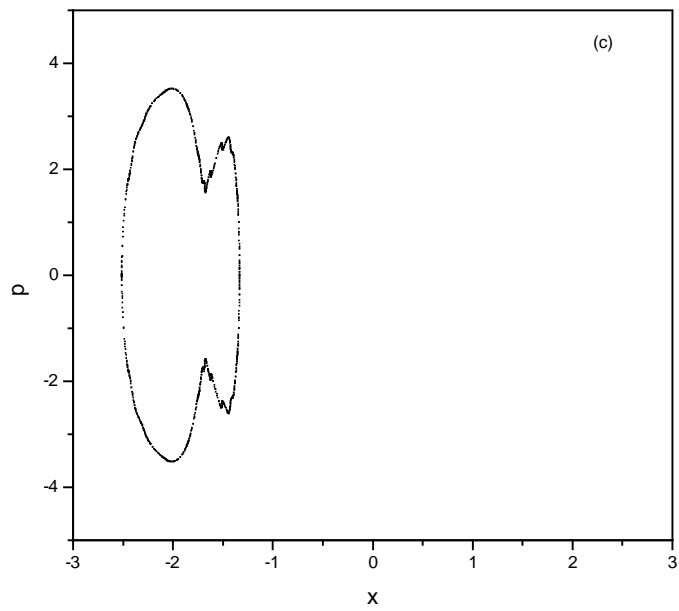
\end{document}